%% file: main.tex
\documentclass[sigconf]{acmart}

\AtBeginDocument{%
  \providecommand\BibTeX{{%
    \normalfont B\kern-0.5em{\scshape i\kern-0.25em b}\kern-0.8em\TeX}}}

\acmConference[arXiv]{priprint}{November, 2021}{preprint}
\acmBooktitle{arXiv,
  November, 2021, preprint}
\acmPrice{15.00}
\acmISBN{978-1-4503-XXXX-X/21/06}

\usepackage{amsmath,amsfonts}
\usepackage{algorithmic}
\usepackage{graphicx}
\usepackage{textcomp}
\usepackage{xcolor}
\usepackage{tipa} 

\usepackage{booktabs} 
\usepackage{multicol}
\usepackage{multirow}

\usepackage{dblfloatfix} 

\usepackage[utf8]{inputenc}

\usepackage{listings}

\usepackage[english]{babel}
\usepackage{blindtext}

\usepackage{wrapfig}

\definecolor{codegreen}{rgb}{0,0.6,0}
\definecolor{codegray}{rgb}{0.5,0.5,0.5}
\definecolor{codepurple}{rgb}{0.58,0,0.82}
\definecolor{backcolour}{rgb}{0.95,0.95,0.92}

\lstdefinestyle{mystyle}{
    backgroundcolor=\color{backcolour},   
    commentstyle=\color{codepurple},
    morecomment=[l][\color{codegreen}]{//},
    keywordstyle=\color{magenta},
    morekeywords={function}
    numberstyle=\tiny\color{codegray},
    stringstyle=\color{codegreen},
    basicstyle=\ttfamily\footnotesize,
    breakatwhitespace=false,         
    breaklines=false,                 
    captionpos=b,                    
    keepspaces=false,                 
    numbers=left,                    
    numbersep=2pt,  
    xleftmargin=1em,
    showspaces=false,                
    showstringspaces=false,
    showtabs=false,                  
    tabsize=1
}

\usepackage{soul}

\usepackage{pifont}

\usepackage[symbol]{footmisc}

\renewcommand\footnotetextcopyrightpermission[1]{}

\begin{document}

\title{Copernicus: Characterizing the Performance Implications of Compression Formats Used in Sparse Workloads}


\author{Bahar Asgari}
\affiliation{%
  \institution{Georgia Institute of Technology}}
\email{bahar.asgari@gatech.edu}

\author{Ramyad Hadidi}
\affiliation{%
  \institution{Georgia Institute of Technology}}
\email{rhadidi@gatech.edu}

\author{Joshua Dierberger}
\affiliation{%
 \institution{Georgia Institute of Technology}}
\email{jdierberger3@gatech.edu}

\author{Charlotte Steinichen}
\affiliation{%
  \institution{Georgia Institute of Technology}}
\email{csteinichen3@gatech.edu}

\author{Amaan Marfatia }
\affiliation{%
  \institution{Georgia Institute of Technology}}
\email{amaan@gatech.edu}

\author{Hyesoon Kim}
\affiliation{%
  \institution{Georgia Institute of Technology}}
\email{hyesoon@cc.gatech.edu}



\renewcommand{\shortauthors}{asgari, et al.}

\begin{abstract}
\input{TEX/abstract.tex}
\end{abstract}


\maketitle

\section{Introduction \& Motivation}
\input{TEX/intro.tex}

\section{Sparse Formats}
\input{TEX/formats.tex}

\section{Sparse Workloads}
\input{TEX/workloads.tex}

\section{Experimental Setup}
\input{TEX/setup.tex}

\section{Copernicus}
\input{TEX/copernicus.tex}

\section{Characterization}
\input{TEX/results.tex}

\section{Related Work}
\input{TEX/related.tex}

\section{Insights \& Conclusions}

\input{TEX/insights.tex}



\bibliographystyle{unsrt}
\balance
\bibliography{ref.bib}

\end{document}

%% file: TEX/abstract.tex
Sparse matrices are the key ingredients of several application domains, from scientific computation to machine learning. The primary challenge with sparse matrices has been efficiently storing and transferring data, for which many sparse formats have been proposed to significantly eliminate zero entries. Such formats, essentially designed to optimize memory footprint, may not be as successful in performing faster processing. In other words, although they allow faster data transfer and improve memory bandwidth utilization – the classic challenge of sparse problems -- their decompression mechanism can potentially create a \textit{computation} bottleneck. Not only is this challenge not resolved, but also it becomes more serious with the advent of domain-specific architectures (DSAs), as they intend to more aggressively improve performance. The performance implications of using various formats along with DSAs, however, has not been extensively studied by prior work. To fill this gap of knowledge, we characterize the impact of using seven frequently used sparse formats on performance, based on a DSA for sparse matrix-vector multiplication (SpMV), implemented on an FPGA using high-level synthesis (HLS) tools, a growing and popular method for developing DSAs. Seeking a fair comparison, we tailor and optimize the HLS implementation of \textit{decompression} for each format. We thoroughly explore diverse metrics, including decompression overhead, latency, balance ratio, throughput, memory bandwidth utilization, resource utilization, and power consumption, on a variety of real-world and synthetic sparse workloads.

%% file: TEX/intro.tex
\label{sec:intro}

Sparse matrices, first distinguished as the main workloads in scientific computation, have become an essential component in many other computation domains such as neural networks, recommendation systems, and graph analytics. Since the primary issue with sparse matrices is storing the enormous amount of non-necessary zero elements, several compression formats have been proposed to efficiently store sparse matrices. While some formats target generality, others are tailored for particular patterns of sparseness (e.g., diagonal matrices) to be more effective in saving them with a minimum storage overhead. Such optimizations for sparse problems mainly focus only on the \textit{storage} overhead in isolation without involving other essential performance metrics such as latency, throughput, and power efficiency. That said, a slow decompression (relative to data transfer) can even surpass the overhead of processing all zero entries in the original dense matrix format; thus, sparse formats may not necessarily guarantee fast execution. This occurs because common sparse formats are often tailored to \textit{the distribution of data}, not \textit{the underlying mechanism of computation}.

The challenges associated with using sparse formats not only are not resolved with the advent of domain-specific architectures (DSAs) for sparse problems, but also gain more importance. DSAs seem to soon become the main platform of sparse computations by approaching the end of Moore's law, proven by the tremendous number of recent studies~\cite{lin2013design, zhu2013accelerating, asgari2019eridanus, mishra2017fine, nurvitadhi2015sparse, gupta2019masr, asgari2020alrescha, pal2018outerspace, hegde2019extensor, kanellopoulos2019smash, dadu2019towards}, to more efficiently accelerate the execution of sparse problems. Prior studies~\cite{gupta2019masr, evansjpeg, jain2018gist, han2016eie, parashar2017scnn, han2017ese, gondimalla2019sparten} have also demonstrated the importance of fast compression/decompression in on-demand applications such as in the inference of neural networks. Regardless of this ongoing research, no study has shed light on the performance implications of using the variety of sparse formats. Particularly, even though prior work has studied the performance implications of software implementations of sparse formats~\cite{blelloch1993segmented, guo2016hybrid, malossi2014performance, goumas2008understanding, karakasis2011exploring, williams2007optimization, sedaghati2015automatic}, the \textit{hardware} implementation of these formats on FPGAs has not been extensively characterized. 

\begin{figure*}[t]
\centering
\vspace{-10pt}
\includegraphics[width=0.95\textwidth]{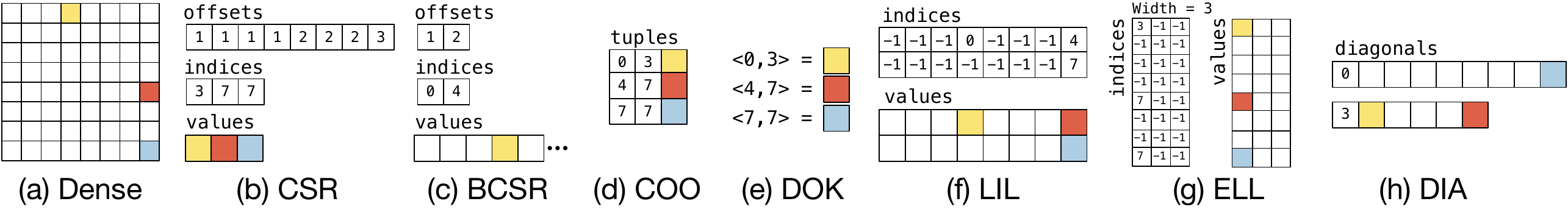}
\vspace{-15pt}
\caption{\textit{Sparse formats: (a)} The original sparse matrix in \textbf{dense} format, \textbf{(b)} \textbf{CSR} including offsets to indicate the number of non-zero entries per row, indices, indicating column indices of non-zero entries, and the values itself. CSC follows the same rule as CSR; \textbf{(c)} \textbf{BCSR} including offsets to indicate the number of non-zero 4$\times$4 blocks per row, indices, indicating the index of the first column of non-zero blocks, and the values, indicating the flatten values of non-zero blocks; \textbf{(d)} \textbf{COO} including a series of (row, column, value) tuples for non-zero values; \textbf{(e)} \textbf{DOK}, which is similar to COO; \textbf{(f)} \textbf{LIL}, which pushes all the non-zero entries to top and saves the row indices; \textbf{(g)} \textbf{ELL}, which is similar to LIL but  pushes the non-zero entries to left ans also uses a padding; and \textbf{(h)} \textbf{DIA}, which saves the non-zero diagonals by adding the diagonal numbers as a header to each diagonal.}
\vspace{-10pt}
\label{fig:formats}
\end{figure*}

To design efficient hardware for processing sparse problems, understanding the impact of the \textit{decompression} mechanism on the performance (measured by various metrics) is crucial. This paper endeavors to address this knowledge gap by performing a thorough \textbf{c}haracterization \textbf{o}n \textbf{per}forma\textbf{n}ce \textbf{i}mplications of \textbf{c}ompression formats \textbf{u}sed in \textbf{s}parse workloads (Copernicus\footnote[2]{Copernicus (/ko\textupsilon 'p\textrevepsilon \textlengthmark rn\textsci k\textipa{@}s/) is a binary star system including five planets.}). We study the diverse metrics, including decompression overhead, latency, throughput, memory bandwidth utilization, resource utilization, and power consumption, of seven frequently used sparse formats on a variety of real-world and synthetic sparse workloads using a hardware platform implemented on a Xilinx FPGA using Vivado high-level synthesis (HLS), a growing and popular method for designing DSAs and quickly prototyping them on FPGAs. We stream compressed sparse data from memory and process them. In an ideal (fast and power-efficient) case, we expect processing to be done at the same pace as receiving data. However, the decompression mechanism can disturb the balance. To fairly evaluate and compare the different decompression mechanisms, for each format, we tailor and optimize the HLS implementation of the decompression. Our suggested HLS implementations and the related performance implications are also applicable to hardware accelerators that directly perform computations on compressed data because they must also reconstruct the location of each non-zero entry.

In summary, Copernicus makes the following contributions:
\begin{itemize}
    \item It is the first work that investigates sparse formats from the \textit{hardware point of view} for accelerating sparse problems.
    \item It thoroughly characterizes diverse metrics for seven frequently used sparse formats, applied on \textit{different partition sizes} of matrices from on-demand applications.
    \item It provides hints to architects to mindfully choose appropriate sparse formats and, more importantly, indicates which parameters must be tuned in ideal hardware to achieve the desired performance or to optimize for a particular metric.
    \item It suggests an optimized pipeline architecture for processing sparse matrices and an HLS-based implementation on an FPGA that can be used as a building block in further accelerators for sparse problems.  
\end{itemize}

%% file: TEX/formats.tex
\label{sec:formats}

\noindent \textit{\textbf{Compressed Sparse Row/Column (CSR/CSC):}}
The CSR/CSC sparse format sequentially stores values in row/column order in a \texttt{values} array while similarly storing their column-index/row-index in a \texttt{indices} array. Another array, \texttt{offsets}, stores index pointers or range for constructing rows/columns. To do so, the adjacent pair of this array [start:stop] represents a slice from the two first arrays. Figure~\ref{fig:formats}b shows an example of CSR. For an $n \times n$ matrix, the length of \texttt{offsets} is $n$ (usually $n+1$, but the first element can store absolute value to reduce the size) and the \textit{maximum}\footnote[2]{Note that these worst-case scenarios are used for on-chip memory allocation. The \textit{storage} overhead is still defined by the number of non-zero entries.} length of \texttt{values} and \texttt{indices} is $n^2$.

\noindent \textit{\textbf{Block CSR/CSC (BCSR/BCSC):}}
The block(-wise) compressed sparse row/column (BCSR/BCSC)~\cite{vuduc2005fast} sparse format is similar to CSR/CSC, but arrays are stored based on the same-shaped blocks (sub-matrices) rather than on the original matrix. This allows block-wise formats to better deal with large matrices. Figure~\ref{fig:formats}c illustrates an example of BCSR for block sizes of 4$\times$4, the block size we choose in all our experiments as well. For an $n \times n$ matrix and $b \times b$ blocks, the length of \texttt{offsets} is $n/b$ and the \textit{maximum} length of \texttt{values} and \texttt{indices} are $n^2$ and  $(n/b)^2$.

\noindent \textit{\textbf{Coordinate (COO) \& Dictionary of Keys (DOK):}}
The COO sparse format simply stores a series of \texttt{tuples}, including the row index, column index, and value for each of the non-zero entries. For an $n \times n$ matrix, the \textit{maximum} length of \texttt{tuples} is $3n^2$.
The DOK format is similar to the COO format except that it stores coordinate-data information as key-value pairs. DOK uses hash tables to store a value with the key of (row index, column index). Figure~\ref{fig:formats}d and e depict an example of COO and DOK, respectively.

\noindent \textit{\textbf{List of List or Linked list (LIL):}}
The LIL~\cite{lil} sparse format stores one list of non-zero elements per row/column. Each element in the lists stores the column/row indices of that row/column, \texttt{indices}, and their value, \texttt{values}. Figure~\ref{fig:formats}f presents an example LIL, which compresses the rows and preserves the columns (this is our assumption for LIL in Copernicus). For an $n \times n$ matrix, the \textit{maximum} length of \texttt{values} and \texttt{indices} is $n$, with $n$ list in total.

\noindent \textit{\textbf{Ellpack (ELL) \& Sliced ELL (SELL):}}
In the ELL~\cite{kincaid1989itpackv} sparse format, non-zero elements are extracted similarly to those of the LIL format, with their column indices and their values. However, they are stored in column-major format with the addition of explicit zero paddings to hold the data for the longest row. This format is ideal for SIMD units since the widths of all \texttt{values} and \texttt{indices} are the same. A sliced ELL (SELL) sparse format first slices the dense matrix row-wise in chunks, and then applies ELL on each chunk. Hence, it reduces the overhead of zero paddings for larger matrices. Figure~\ref{fig:formats}g shows an example of ELL with a padding width of three. In Copernicus, we set this width to six. For an $n \times n$ matrix, the \textit{maximum} length of \texttt{values} and \texttt{indices} is $n$ (longest possible row). The width in ELL is $n$, and that in SELL varies based on the pattern of data. Variants of ELL formats such as ELL+COO, Jagged Diagonal Storage (JDS)~\cite{saad1989numerical}, and SELL-C-$\sigma$~\cite{kreutzer2014unified} are also popular. ELL+COO mixes ELL and COO formats to reduce the width of long rows. The JDS format sorts the rows in ELL from longest to shortest (for vector machines). SELL-C-$\sigma$ is a variant of JDS that only sorts rows within a window of $\sigma$.

\noindent \textit{\textbf{Diagonal (DIA):}}
The DIA~\cite{saad2003iterative} sparse format operates by specifying a diagonal number (0 for the main diagonal, negative/positive for diagonals which start on a lower/higher row/column) followed by the values that fall on the diagonal, \texttt{diagonals}. Figure~\ref{fig:formats}h illustrates an example of DIA. For an $n \times n$ matrix, the \textit{maximum} number of non-zero diagonals is $2n-1$ and the \textit{maximum} length of a diagonal is $n+1$ (the additional element contains the diagonal number). 

%% file: TEX/workloads.tex
\label{sec:workloads}

Sparse matrices are the main data structure in several application domains, from scientific computations to graph analytics. This section first introduces various instances of sparse matrices and lists our \textit{real-world} workloads. Then, it reviews the common regular structures of sparseness and introduces our \textit{synthetic} workloads to evaluate a wider range of sparse applications. Finally, this section explains the common computation of several sparse problems, on which we build our hardware platform for evaluation.     

\subsection{Sparse Matrices in Various Domains}

The main ingredient of different fields of \textit{scientific computations} is modeling physical phenomena, such as sound, heat, elasticity, fluid dynamics, and quantum mechanics. Such models often use partial differential equations (PDEs). To use digital computers for solving PDEs, they are discretized into a 3D grid. Discretization is used to convert PDEs to a linear system of algebraic equations, Ax = b, in which A is the coefficient matrix. Since not all the points in a 3D grid are used in the discretization of a phenomenon, \textit{the coefficient matrix A is sparse}. Sparse matrices also exist in \textit{graph analytics}. A common approach for representing graphs is to use an adjacency matrix, each entry of which represents an edge in the graph. As in many applications, not all the nodes in a graph are connected; the graphs are sparse and so are their equivalent \textit{adjacency matrice}s containing several zeros. As a representative for various size sparse matrices in the domain of scientific computation and graph analytics, with different density, we obtained the matrices listed in Table~\ref{table:SuiteSparse} from SuiteSparse~\cite{davis2011university} matrix collection.  

The third group of common sparse problems is the applications of machine learning, including the \textit{inference of neural networks} and \textit{the recommendation systems}. Since after training, close-to-zero values are assigned to several model parameters, a common practice is to prune those values. Pruning results in sparse \textit{weight matrices}. Sparse matrices in neural networks are often not as sparse as extremely sparse matrices in the first two application domains. Besides, in neural networks, sparseness is more random and varies case by case with the algorithm of pruning and the other design factors such as the desired accuracy. The recommendation system models are the other instance of sparse problems. They include an embedding table followed by a neural network. Although the embedding tables are dense, accesses to them are random and sparse~\cite{naumov2019deep}. To evaluate the whole range of the variety of randomness in sparse machine learning applications, Copernicus covers the synthetic matrices introduced in the following.

\renewcommand{\arraystretch}{0.9}
\begin {table}[t]
\small
\begin{center} 
\vspace{-0pt}
\caption{Matrices from SuiteSparse~\cite{davis2011university} matrix collection.}{
\vspace{-12pt}
\resizebox{1\columnwidth}{!}{
\begin{tabular}{ c c c c c }
              \toprule
              \textbf{ID} & \textbf{Name} & \textbf{Dim.(M)$^1$} & \textbf{NNZ(M)$^2$} &  \textbf{Kind} \\
              \midrule
              2C & 2cubes\_sphere & 0.101 & 1.647 & Electromagnetics Problem\\
              \midrule
              FR & Freescale2 & 2.9 & 14.3 & Circuit Sim. Matrix \\ 
              \midrule
              RE & N\_reactome & 0.016 & 0.043 & Biochemical Network \\ 
              \midrule
              AM & amazon0601 & 0.4 & 3.3 & Directed Graph \\ 
              \midrule
              DW & dwt\_918 & 0.000918 & 0.0073 & Structural Problem\\ 
              \midrule
              EO & europe\_osm & 50.9 & 108 & Undirected Graph \\ 
              \midrule
              FL & flickr & 0.82 & 9.8 & Directed Graph \\ 
              \midrule
              HC & hcircuit & 0.1 & 0.51 & Circuit Sim. Problem \\ 
              \midrule
              HU & hugebubbles & 18.3 & 54.9 & Undirected Graph \\ 
              \midrule
              KR & kron\_g500-logn21 & 2 & 182 & Undirected Multigraph \\ 
              \midrule
              RL & rail582 & 0.056 & 0.4 & Linear Prog. Problem \\ 
              \midrule
              RJ & rajat31 & 4.6 & 20.3 & Circuit Sim. Problem \\ 
              \midrule
              RO & roadNet-TX & 1.3 & 3.8 & Undirected Graph \\ 
              \midrule
              RC & road\_central & 14 & 33.8 & Undirected Graph \\ 
              \midrule
              LJ & soc-LiveJournal1 & 4.8 & 68.9 & Directed Graph \\ 
              \midrule
              TH & thermomech\_dK & 0.2 & 2.8 & Thermal Problem \\ 
              \midrule
              WE & wb-edu & 9.8 & 57.1 & Directed Graph \\ 
              \midrule
              WG & web-Google & 0.91 & 5.1 & Directed Graph \\ 
              \midrule
              WT & wiki-Talk & 2.3 & 5 & Directed Graph \\ 
              \midrule
              WI & wikipedia & 3.5 & 45 & Directed Graph \\ 
              \midrule 
              \multicolumn{5}{l}{\small{$^1$ Dim.: dimension or the number of columns/rows of a square matrix.}}\\
              \multicolumn{5}{l}{\small{$^2$ NNZ: the number of non-zero entries.}}\\
\end{tabular}}}
\label{table:SuiteSparse}
\end{center}
\vspace{-15pt}
\end{table}
\renewcommand{\arraystretch}{1.0}

\subsection{Synthetic Sparse Matrices}
\label{sec:synthetic}
Besides the real-world matrices listed in Table~\ref{table:SuiteSparse}, our workloads consist of two groups of synthetic sparse matrices. The first group includes \textit{randomly generated} sparse matrices, the density of which varies from 0.0001 to 0.5. We generate the denser random matrices (i.e., the density of 0.1 to 0.5) as a representation for those in machine learning applications. On the other hand, the more sparse random matrices (i.e., density between 0.0001 to 0.01) represent scientific and graph applications with no particular structure. The second group of our synthetic sparse matrices denotes the common structure in sparse matrices: \textit{diagonal and band matrices}. A band matrix is a sparse matrix, the non-zero entries of which are confined to a diagonal band, including the main diagonal and more than one diagonal on each side. The width of a band matrix is the number $k$ such that $a_{i,j}=0$ if $|i-j| > k/2$. We generate and evaluate band matrices of size 8000 with widths of 2, 4, 16, 32, and 64. Numerical problems in higher dimensions often lead to band matrices (e.g., a PDE on a square domain). A type of band matrices consisting of only the main diagonal (i.e., $k=1$) is called a diagonal matrix. Diagonal matrices also occur in many fields of linear algebra.

\subsection{Common Computation in Sparse Problems}

\noindent This section shows that sparse matrix-vector multiplication (SpMV) is the key sparse kernel in all of the three aforementioned domains of sparse problems. Designing DSAs for accelerating SpMV has been the focus of several recent studies~\cite{qin2020sigma, zhang2020sparch, srivastava2020tensaurus, gondimalla2019sparten, kanellopoulos2019smash, sadi2019efficient, hegde2019extensor, asgari2020alrescha} 

\textit{\textbf{Scientific Computations:}}
The coefficient sparse matrix $A$ in a linear system of algebraic equations is very large for two or higher dimensional problems (e.g., elliptic, parabolic, or hyperbolic PDEs). While for small $A$s, solving  PDEs is doable by direct methods -- those based on matrix inversion -- they become less efficient for larger $A$s. Such systems of linear equations with a large symmetric positive-definite matrix $A$ can be solved by iterative algorithms such as conjugate gradient (CG) methods. In both cases (i.e., direct and iterative methods) for solving PDEs, the key sparse kernel is \textit{SpMV}. Symmetric Gauss-Seidel iteration~\cite{golub2012matrix} used in the CG algorithm or the lower-upper (LU) decomposition used for matrix inversion are examples that contain \textit{SpMV}. 

\textit{\textbf{Graph Analytics:}}
Graph algorithms, such as breadth-first search, single-source shortest path, and PageRank, which traverse vertices and edges of a graph to compute some properties based on the graph relationships, can be implemented as a sparse matrix-vector operation. Such an implementation is compatible with the vertex-centric programming model~\cite{malewicz2010pregel}, which divides a graph algorithm into two main phases. In the first phase, a row of the adjacency matrix is processed using a vector-vector operation between the row of the matrix, and a property vector varied based on the algorithm type. Then, the output vector from the first phase is reduced by some form of reduction operations (e.g., summation). The combination of the two phases creates a dot-product, and combining the dot-products for processing the whole graph leads to an \textit{SpMV}. 

\textit{\textbf{Machine Learning:}}
Machine learning applications consist of SpMV or sparse matrix-matrix multiplication, both of which rely on the same underlying dot-product engine. For instance, in convolution neural networks, the computation of convolutional layers can be viewed as a matrix-matrix multiplication. In other words, convolving a 3D input with a given number of filters can be represented as an equivalent matrix-matrix multiplication that multiplies the 2D flatten weight matrix by the input matrix, which is originally two-dimensional. In recommendation systems, the sparse embedding-table look-ups end up as a reduction operation (e.g., summation) that can also be implemented using a dot-product engine, initially designed for \textit{SpMV}.


%% file: TEX/setup.tex
\label{sec:setup}

\subsection{Platform and System Overview}

We describe our hardware platform in C++ and use this implementation for generating RTL to be executed on an FPGA. To generate RTL (in Verilog), we use the high-level synthesis (HLS) tool of Vivado and related \texttt{\#pragma}s as hints to detail the architecture. After simulations in Vivado\_HLS, we use Vivado to synthesise and implement our hardware on an xq7z020 FPGA from Zynq-7000 family connected to a DDR3 memory. The clock frequency is set to 250\,MHz. We verify the functionality of the RTL using synthetic data as a C++ testbench for C/RTL co-simulation. The inputs to our hardware platform are sparse workloads (SuitSparse and synthetic sparse matrices). We use Matlab to preprocess the workloads and compress them in target formats, studied in Copernicus. 

Although all sparse formats eliminate transferring a large portion of zero entries, applying them directly on the large original matrices is not efficient. For instance, formats such as CSR transfer one index for all rows, even for all-zero rows. Therefore, their overhead grows with the size (i.e., dimension) of the matrix. Therefore, as compressing and transferring large units of data (e.g., the entire original matrix) is not beneficial for some of the formats, a common efficient practice is to apply the compression on the smaller partitions of the original matrix, which also allows coarse-grained parallel processing of partitions. Additionally, since some sparse matrices in real-world applications capture a certain level of spatial locality, by using partitioning, we can eliminate transferring and processing the all-zero partitions. Following this common practice, Copernicus also applies all the compression formats \textit{only on the non-zero partitions} of large matrices. The size of these partitions is one of our hyperparameters.

\subsection{Metrics \& Hyperparameters}

Copernicus evaluates the performance implications of sparse formats using the metrics introduced in the following. First, we define $\sigma$ as a metric to measure the latency \textbf{overhead of decompression}:
\begin{equation}
\small
\vspace{-0pt}
  \sigma = \frac{T_{decomp} + nnz\_rows \times T_{dot}}{p \times T_{dot}}
  \label{equ:sigma}
\vspace{-0pt}
\end{equation}

\noindent in which $T_{decomp}$ indicates decompression latency, which consists of latency for BRAM accesses and logic, $nnz\_rows$ are non-zero rows for which we must perform a dot-product, each taking $T_{dot}$, and $p$ is the partition size. As a result, for the dense format, $\sigma =1$. Besides $\sigma$, we measure the breakdowns of latency: (i) \textbf{memory latency}, the time to transfer a compressed partition (data and metadata) to FPGA and buffer it in the BRAM; and (ii) \textbf{computation latency} consisting of decompression, dot-product, and necessary BRAM accesses. Furthermore, seeking an appropriate sparse format for achieving balanced streaming, we proposed evaluating a \textbf{balance ratio}, which we define as the average ratio of memory latency to compute latency for all non-zero partitions. The balanced ratio of perfectly balanced streaming would be one. An imbalance streaming leads to idle computation or pauses in data transfer. 

As another important factor in the execution of sparse problems, we evaluate \textbf{throughput}, defined as bytes processed per second, which reflects the bubbles in the streaming pipeline caused by imbalance streaming (balance ratio $\neq$ 1). Besides throughput, we compare the \textbf{memory-bandwidth utilization}, the ratio of useful data over all transmitted data (i.e., useful data plus metadata)
Our other metrics for the full design-space exploration are \textbf{resource utilization} and dynamic/static \textbf{power consumption}. By resource utilization, we mean the percentage of FPGA resources used by all components (entire Figure~\ref{fig:micro}). We evaluate the aforementioned metrics while varying the hyperparameters including practical partition sizes of 8, 16, and 32, and the width of 2, 4, 8, 16, 32, and 64 for band matrices to represent realistic matrices, and the density of random matrices from 0.0001 to 0.5.

%% file: TEX/copernicus.tex
\label{sec:copernicus}

\subsection{Architecture}

\begin{figure}[t]
\centering
\vspace{-0pt}
\includegraphics[width=0.49\textwidth]{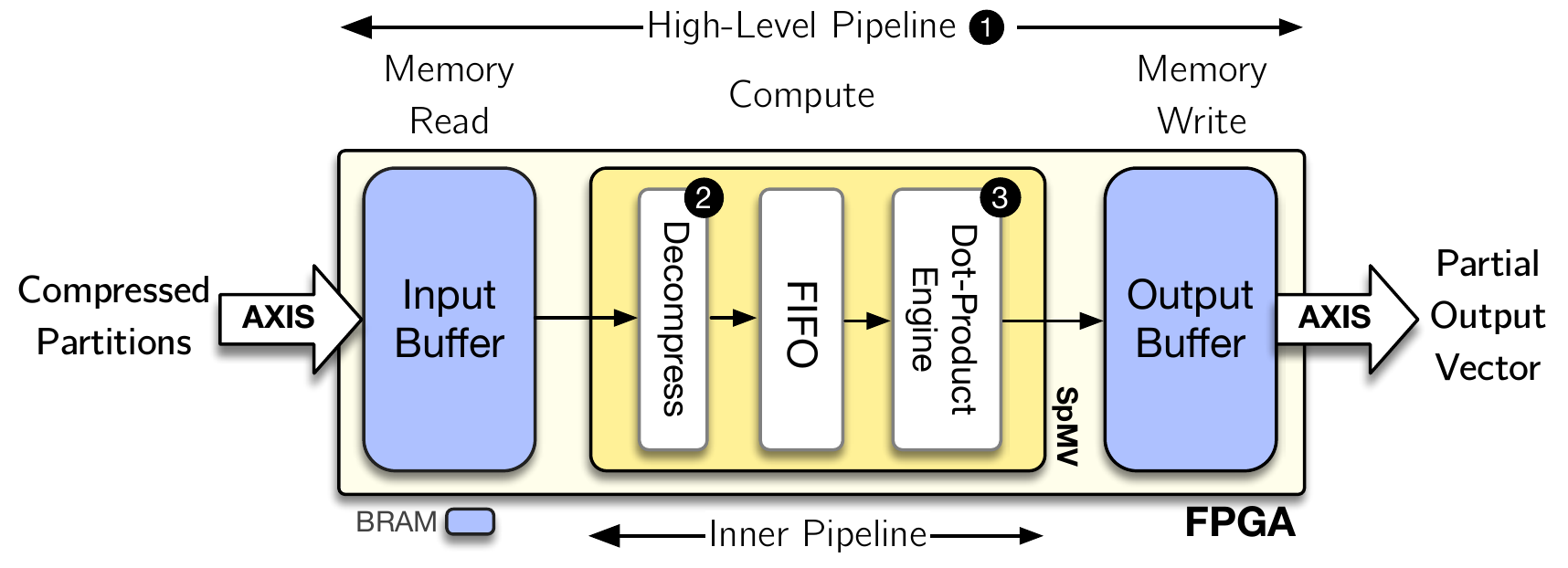}
\vspace{-20pt}
\caption{ \textit{The architecture of our evaluation platform:} Streaming the compressed partitions of a sparse matrix from the memory to FPGA through AXI stream interfaces and processing them (i.e., SpMV) in a pipeline. The \textit{decompression} component varies based on sparse format.}
\vspace{-18pt}
\label{fig:micro}
\end{figure}

This section proposes our hardware architecture for processing sparse matrices (i.e., executing SpMV). To be effective in accelerating any problem, two ingredients are key: parallel computation and fast accesses to memory. Parallelizing the computation of SpMV is relatively straightforward and can be implemented in various ways. For instance, the vector operand can be multiplied with the non-zero rows of the sparse matrix operand in parallel, or the columns of the matrix and the vector operand can be partitioned into a few chunks and be processed concurrently. On top of these coarse-grained approaches, element-wise fine-grained parallelism can always be applied to each of the dot products, the fundamental operations in SpMV. For instance, the intuitive way to perform parallel dot products is by implementing a fixed-size compute logic (i.e., a multiplier array attached to an adder tree) and using it on the chunks of the operands of the original SpMV. We build our hardware platform (shown in Figure~\ref{fig:micro}) based on the building blocks of such a fine-grained-parallel dot product engine (Figure~\ref{fig:micro}, \ding{184}) and investigate how to effectively integrate it with stream memory accesses as the second fundamental ingredient to achieve high performance. Instances of this architecture can be aggregated for implementing coarse-grain parallelism.

The components of our evaluation architecture to stream the compressed partitions and process them in an FPGA include (i) the global memory, (ii) an AXI streaming (AXIS) interface through which the partitions and the vector operand of SpMV are transferred from the memory to FPGA, and (iii) the high-level three-stage pipeline (Figure~\ref{fig:micro}, \ding{182}) implemented in the FPGA that receives the partitions in the memory-read stage, processes them in the compute stage (i.e., SpMV block in Figure~\ref{fig:micro}), and streams the partial output vector back to the memory in the memory-write stage. The input buffer contains a  partition compressed in a particular format (e.g., it contains values, offsets, and column indices for CSR). The following explains the details of the compute stage of this high-level pipeline, which itself comprises a two-stage pipeline, including decompress and dot-product stages. The block of SpMV iteratively creates dense non-zero rows in the first stage (Figure~\ref{fig:micro}, \ding{183}). Then, the second stage (Figure~\ref{fig:micro}, \ding{184}) performs a dot-product between the result of the first stage and the vector operand of SpMV. We implement the dot-product as an array of multipliers connected to a balanced adder tree. Since the output of the SpMV is a vector (not necessarily sparse), we do not include a recompression stage in hardware. 

\begin{figure}[t]
\centering
\vspace{-0pt}
\includegraphics[width=0.45\textwidth]{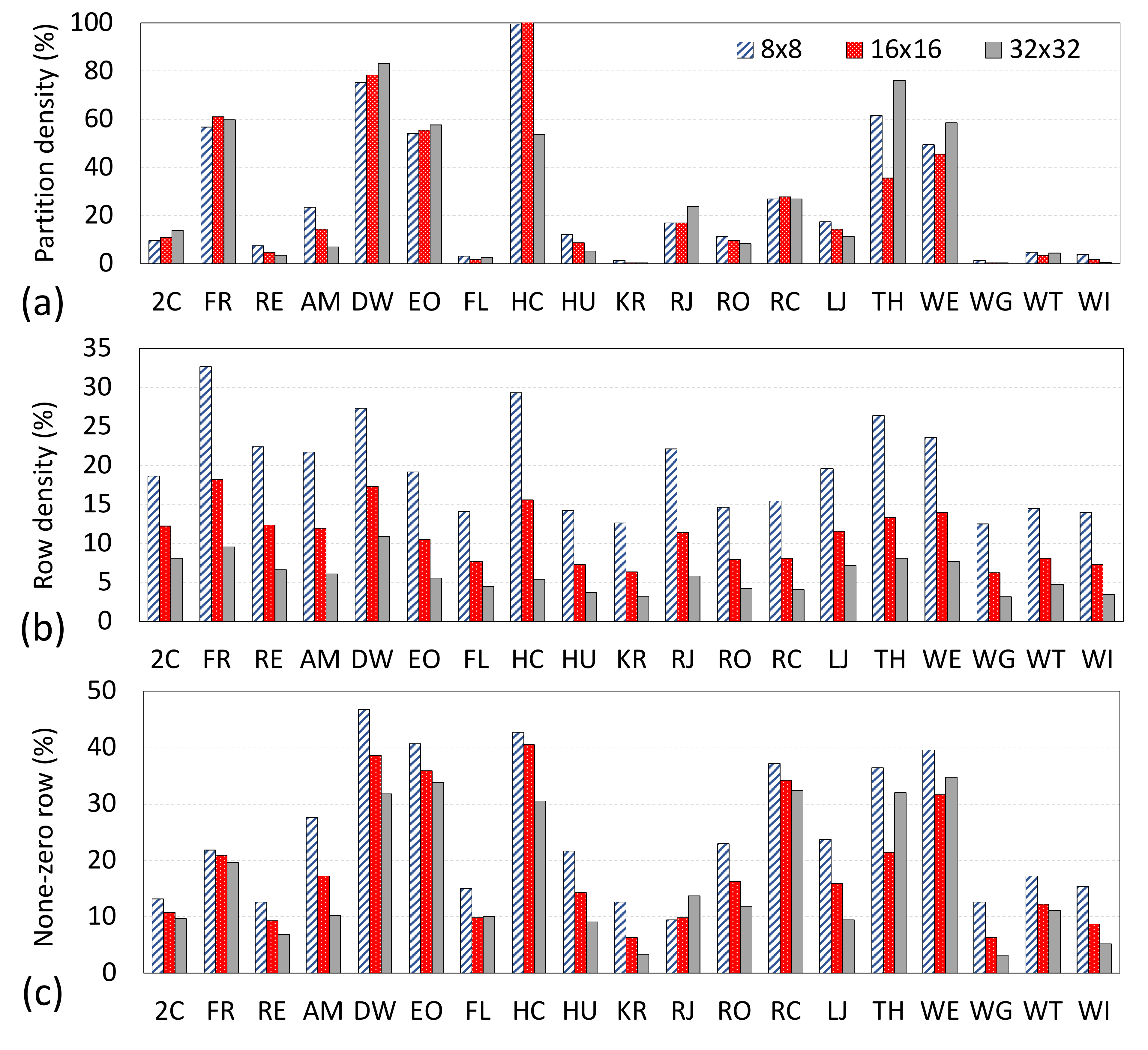}
\vspace{-15pt}
\caption{\textit{Density and spatial locality:} the average percentage of (a) non-zero values in partitions, (b) non-zero values in non-zero rows, and (c) non-zero rows in partitions for the sparse matrices listed in Table~\ref{table:SuiteSparse}.}
\vspace{-15pt}
\label{fig:density}
\end{figure}

We use this architecture (Figure~\ref{fig:micro}) as the baseline and tailor the decompression stage for each format (Section~\ref{sec:decomp}) and then compare the outcomes. The \textit{decompression} (Figure~\ref{fig:micro}, \ding{183}) is the bottleneck-prone stage and the focus of Copernicus. In our platform, the width of the dot-product engine (Figure~\ref{fig:micro}, \ding{184}) is the same as the width of the partitions. The \textit{partition density} and, more specifically the \textit{row density} (i.e., the density of non-zero rows), defines the computation utilization of the dot-product engine at run time. Besides, the \textit{number of non-zero rows} in the partitions determines the utilization of the inner pipeline. Figure~\ref{fig:density} illustrates the three mentioned factors for three partition sizes: 8, 16, and 32. These raw statistics can be interpreted along with the evaluation results reported in Section~\ref{sec:copernicus}.

\subsection{Optimized Decompression in HLS}
\label{sec:decomp}

\noindent Here, we discuss our tailored HLS implementation of the decompression mechanism (Figure~\ref{fig:micro}, \ding{183}) for the seven target sparse formats.

\textit{\textbf{CSR:}}
As Listing~\ref{code:csr} shows, since CSR uses three vectors (i.e., offsets, column indices, and values) to represent a sparse matrix, for decompressing a non-zero row, we need to first access the offsets (Listing~\ref{code:csr}, line 7), and then, based on \texttt{numVal}, we can read as many column indices and values as required (Listing~\ref{code:csr}, line 10). As a result, decompression from the CSR format is likely to be compute-bound because of the overhead of one extra access to BRAM. Additionally, to retrieve the column indices and values, \textit{since accesses to the BRAM blocks are sequential}, we do not know in advance which elements of column indices and values are going to be accessed. Thus, we cannot partition and allocate those two vectors across the blocks of BRAM to guarantee parallel accesses. Because of the sequential accesses in a non-zero row, the latency of decompressing a row depends on the number of non-zero elements in that row. By assuming that we stream the offsets and column indices using two streamlines in parallel, the one with more non-zero elements (longer one) defines the latency of memory access. To reduce the negative impact of the accesses to offsets on performance, we pipeline this progress to concurrently create non-zero rows (if more than one).   
\begin{lstlisting}[label=code:csr,caption=CSR-decompression HLS pseudo code, language=Octave, style=mystyle]
function decompressCSR(A, readInx, oldInx)
    // A in CSR: offsets[OFFSET_LENGTH]
    //           colInx[COL_INX_LENGTH]
    //           values[VAL_LENGTH]
    // readInx: current row index
    // oldInx: last row index
    numVal = offsets[readInx] - offsets[readInx-1]
    for i=0 to numVal:
        #pragma HLS pipeline
        drow[colInx[oldInx+i]] = values[oldInx+i]
return drow
\end{lstlisting}

\textit{\textbf{BCSR:}}
As Listing~\ref{code:bcsr} shows, the decompression of BCSR is similar to that of CSR, whereas instead of individual non-zero elements, the non-zero blocks are processed. To initiate the accesses to column indices and values, one extra access to the offsets is required per each row of blocks (Listing~\ref{code:bcsr}, line 9). The advantage of BCSR over CSR is that we can distribute the values and column indices over BRAM blocks and access their elements in parallel, for which, as lines 1 and 2 in Listing~\ref{code:bcsr} show, we completely partition the \texttt{values} and \texttt{colInx} across their second dimension before calling the decompression function. This allows us to unroll the for loop (line 12), the iterations of which access different BRAM blocks in parallel. The downsids of BCSR, however, are (i) the overhead of transferring zero elements in the non-zero blocks and (ii) processing all the rows in the non-zero blocks whether they (the rows) are all zero or not. The latency of decompressing the blocks in a row depends on the number of non-zero \textit{blocks} in that row. On the other hand, since the \texttt{values} has the longest length (compared to \texttt{offsets} and \texttt{colInx}), transferring \texttt{values} defines the memory latency.
\begin{lstlisting}[label=code:bcsr,caption=BCSR-decompression HLS pseudo code, language=Octave, style=mystyle]
#pragma HLS array_partition variable=values dim=2
#pragma HLS array_partition variable=colInx dim=2
function decompressBCSR(A, readInx, oldInx)
  // A in BCSR: offsets[OFFSET_LENGTH]
  //            colInx[COL_INX_LENGTH]
  //            values[VAL_LENGTH][VAL_WIDTH]
  // readInx: current row index
  // oldInx: last row index
  numBlocks = offsets[readInx] - offsets[readInx-1]
  for i=0 to numBlocks:
    for j=0 to VAL_WIDTH:
      #pragma HLS unroll
      drows[j / BLOCK_LENGTH][colInx[oldInx + i] 
      + j mod BLOCK_LENGTH] = values[oldInx + i][j]
return drows
\end{lstlisting}

\textit{\textbf{CSC:}}
Listing~\ref{code:csc} shows the pseudo code of CSC decompression, in which the \textit{columns} are compressed. On the contrary, the hardware requires \textit{rows} of the matrix for performing SpMV. Because of this mismatch, the decompression mechanism must iteratively traverse all the columns of the matrix to find the values corresponding to the current row (Listing~\ref{code:csc}, line 12). Although this mismatch makes the decompression inefficient, we still include this extreme case in our evaluation to explore how much performance is hurt if the format and the hardware are not aligned.  
\begin{lstlisting}[label=code:csc,caption=CSC-decompression HLS pseudo code, language=Octave, style=mystyle]
function decompressCSC(A, readInx)
    // A in CSC: offsets[OFFSET_LENGTH]
    //           rowInx[ROW_INX_LENGTH]
    //           values[VAL_LENGTH]
    // readInx: current row index
    numVal = offsets[colInx] - offsets[Inx-1]
    for i=0 to CSC_ROW_INX_LENGTH 
                && colInx < CSC_OFFSETS_LENGTH:
        	startInx = i
        	#pragma HLS pipeline
        	while i < startInx + numVal:
        	   if rowInx[i] == read_inx:
        	        drow[colInx-1] = values[i]
        	        break
        	   i++
        	i = offsets[colInx++]
return drow
\end{lstlisting}

\textit{\textbf{LIL:}}
The LIL decompression (Listing~\ref{code:lil}) avoids extra accesses to BRAM and enables \textit{deterministic} parallel accesses to the BRAM blocks for decompressing the non-zero rows of a sparse matrix. Since the columns of \texttt{values} and \texttt{indices} can always be accessed in parallel, we partition both of them (Listing~\ref{code:lil}, lines 1 and 2) before calling decompression. As a result, no extra read access is required to determine the number of next read accesses. Thus, the latency of processing a matrix depends on the number of non-zero rows. In other words, creating a non-zero row consists of the latency of one BRAM access (since the accesses are parallel) plus the latency for creating the input of the dot product, which is done by a simpler logic compared to those of CSR or BCSR. To recognize the end of the non-zero rows, one additional BRAM access is required. Memory latency for LIL is defined by the number of non-zero rows, the size of rows, and transferring one additional row for indicating the end of non-zero rows.
\begin{lstlisting}[label=code:lil,caption=LIL-decompression HLS pseudo code, language=Octave, style=mystyle]
#pragma HLS array_partition variable=values dim=2
#pragma HLS array_partition variable=Inx dim=2
function decompressLIL(A, readInx[], oldInx)
  // A in LIL: values[HEIGTH][WIDTH]
  //           Inx[HEIGHT][WIDTH]
  // readInx: current row index
  // oldInx: last row index
  minInx = inf
  for i=0 to WIDTH:
   #pragma HLS pipeline
   if readInx[i]<HEIGTH && Inx[readInx[i]][i]<minInx:
       minInx = Inx[readInx[i]][i]
   for i=0 to WIDTH:
      #pragma HLS unroll
      if Inx[readInx[i]][i] == minInx:
        drow[i] = values[readInx[i]][i]
        readInx[i]++
return drow
\end{lstlisting}

\textit{\textbf{ELL:}}
Similar to LIL, the representation of matrix \texttt{A} in ELL (Listing~\ref{code:ell}) includes \texttt{values} and \texttt{indices} that can be accessed in parallel and thus are partitioned and distributed over BRAM blocks (Listing~\ref{code:ell}, lines 1 and 2), which subsequently allows unrolling the for loop for parallel processing (line 7). The difference between LIL and ELL, however, is the direction of compression. Although the direction of compression in ELL enables a simple assignment shown in line 8, it prevents skipping the all-zero rows that in turn can cause a performance drop. Since we completely unroll the for loop (line 7), reducing \texttt{ELL\_MAX\_COMP\_ROW\_LENGTH} in the ELL implementation and using optimizations such as ELL-COO only impact the resource utilization of FPGA, not the performance.  
\begin{lstlisting}[label=code:ell,caption=ELL-decompression HLS pseudo code, language=Octave, style=mystyle]
#pragma HLS array_partition variable=values dim=2
#pragma HLS array_partition variable=indices dim=2
function decompressELL(A)
    // A in ELL: values[ELL_MAX_COMP_ROW_LENGTH]
    //           indices[ELL_MAX_COMP_ROW_LENGTH]
    for i=0 to ELL_MAX_COMP_ROW_LENGTH:
      #pragma HLS unroll
      drow[indices[i]] = values[i]
return drow
\end{lstlisting}

\begin{figure*}[t]
\centering
\vspace{-15pt}
\includegraphics[width=1\textwidth]{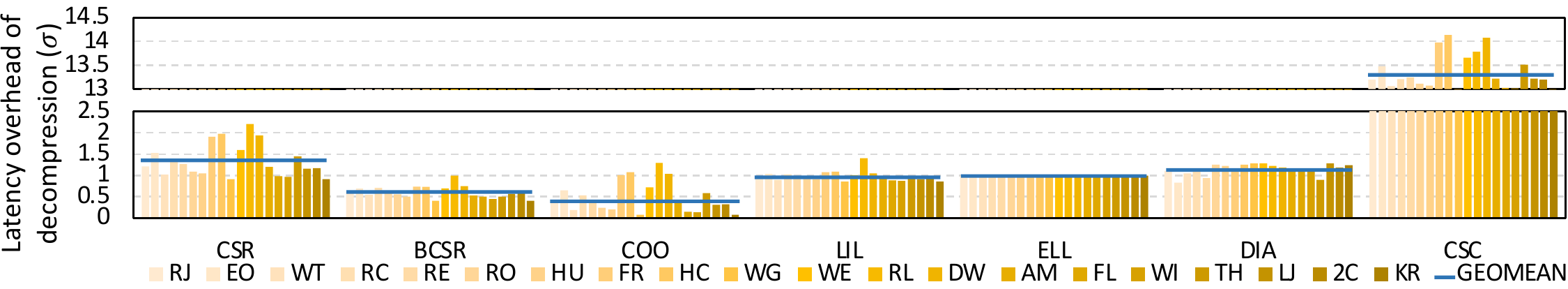}
\vspace{-20pt}
\caption{\textit{Decompression overhead for SuiteSparse:} comparing the latency overhead, $\sigma$ (lower is betters), of seven sparse formats for partition size of $16 \times 16$. A darker color indicates less sparsity (i.e., higher density).}
\vspace{-12pt}
\label{fig:latency_SS}
\end{figure*}

\textit{\textbf{COO:}}
As Listing~\ref{code:coo} shows, since COO saves tuples for representing a matrix, its decompression mechanism is pretty straightforward, including a simple assignment, shown in Listing~\ref{code:coo}, line 7. The downside of COO, however, is that we do not know in advance how many elements exist in each row. Thus, we cannot partition and allocate the vector of \texttt{tuples} across the blocks of BRAM to guarantee parallel accesses. For the same reason, we pipeline the for loop (line 5) rather than unrolling it. The same procedure is also applicable to DOK.
\begin{lstlisting}[label=code:coo,caption=COO-decompression HLS pseudo code, language=Octave, style=mystyle]
function decompressCOO(A, row)
  // A in COO: tuples[COO_NUM_TUPLES][3]
  // row: the current row
  for i=0 to COO_NUM_TUPLES:
    #pragma HLS pipeline
    if (tuples[i][0]!=inf && tuples[i][0]==row):
        drow[tuples[i][1]] = tuples[i][2];
return drow
\end{lstlisting}

\textit{\textbf{DIA:}}
Line 7 of Listing~\ref{code:dia} shows the pseudo code for the decompression mechanism of DIA, the most domain-specific format studied in Copernicus. DIA saves matrix A as \texttt{diags}, a two-dimensional matrix including all non-zero diagonals of matrix A. The first element of each row of \texttt{diags} indicates the diagonal number. To decompress the rows of matrix A, the decompression function traverses all rows of \texttt{diags} to find the elements corresponding to the current \texttt{row}. To this end, we use two helper functions, \texttt{DiaInxForRow} (line 1) and \texttt{IsRowOnDiagonal} (line 4). As the decompression mechanism suggests, although in terms of memory footprint DIA should be beneficial for diagonal matrices, its decompression mechanism is not quite compatible with even a simple computation such as a fine-grained parallel SpMV, which subsequently requires rows of the matrix. Such an overhead worsens when non-zero elements are scattered over multiple diagonals but do not completely fill them.

\begin{lstlisting}[label=code:dia,caption=DIA-decompression HLS pseudo code, language=Octave, style=mystyle]
function DiaInxForRow(row, d)
    return (row + d < row) ? row + d : row

function IsRowOnDiagonal(row, d) 
    return d <= WIDTH - 1 - row && d >= -row

function decompressDIA(A, row)
  // A in DIA: 
  //   diags[NUM_DIAGONALS][MAX_DIAGONAL_LEN]
  // row: the current row
  for i=0 to NUM_DIAGONALS:
    #pragma HLS pipeline
    d = diags[i][0]
  	  if (!IsRowOnDiagonal(row, d)) continue
  	  // column index is the row + d
  	  // + 1 since 0 element is the diagonal number
  	  drow[row+d] = diags[i][DiaInxForRow(row,d)+1]
return drow
\end{lstlisting}

%% file: TEX/results.tex
\subsection{The Overhead of Decompression}
\label{sec:latency}


\begin{figure}[t]
\centering
\vspace{-0pt}
\includegraphics[width=0.45\textwidth]{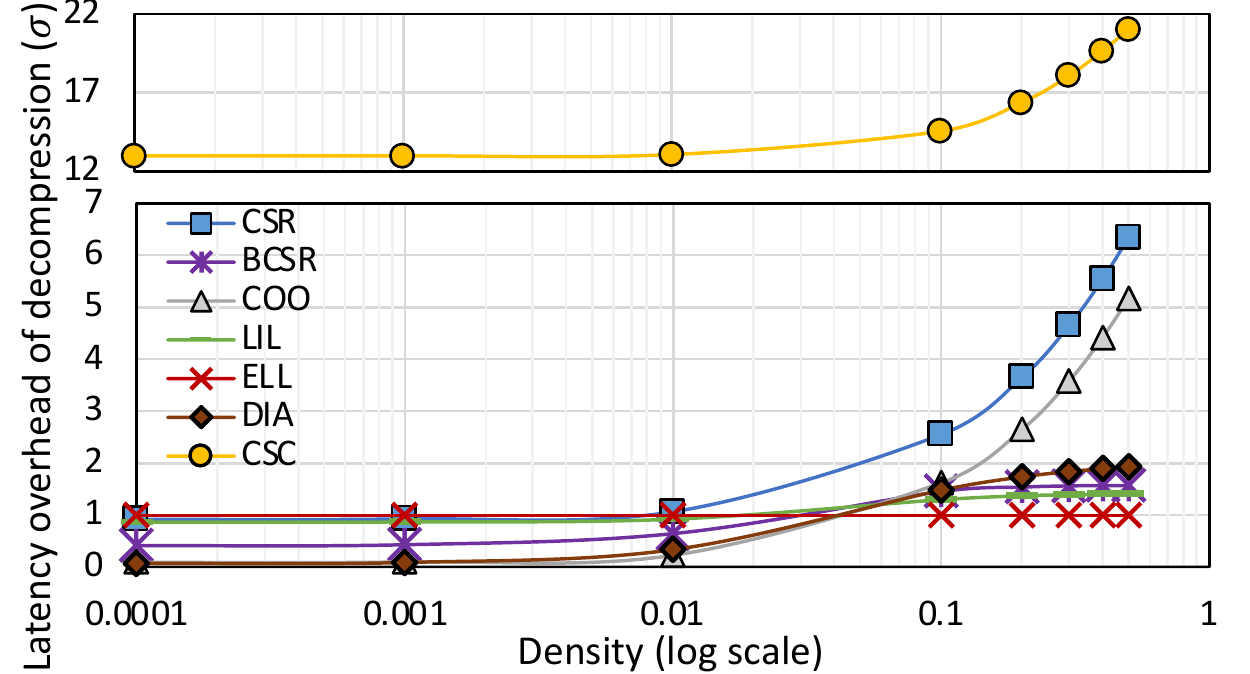}
\vspace{-10pt}
\caption{\textit{Decompression overhead for random matrices:} comparing the $\sigma$ (lower is betters) of seven sparse formats for $16 \times 16$ partitions  when density varies from 0.0001 to 0.5.}
\vspace{-15pt}
\label{fig:latency_rand}
\end{figure}

\begin{figure}[b]
\centering
\vspace{-15pt}
\includegraphics[width=0.45\textwidth]{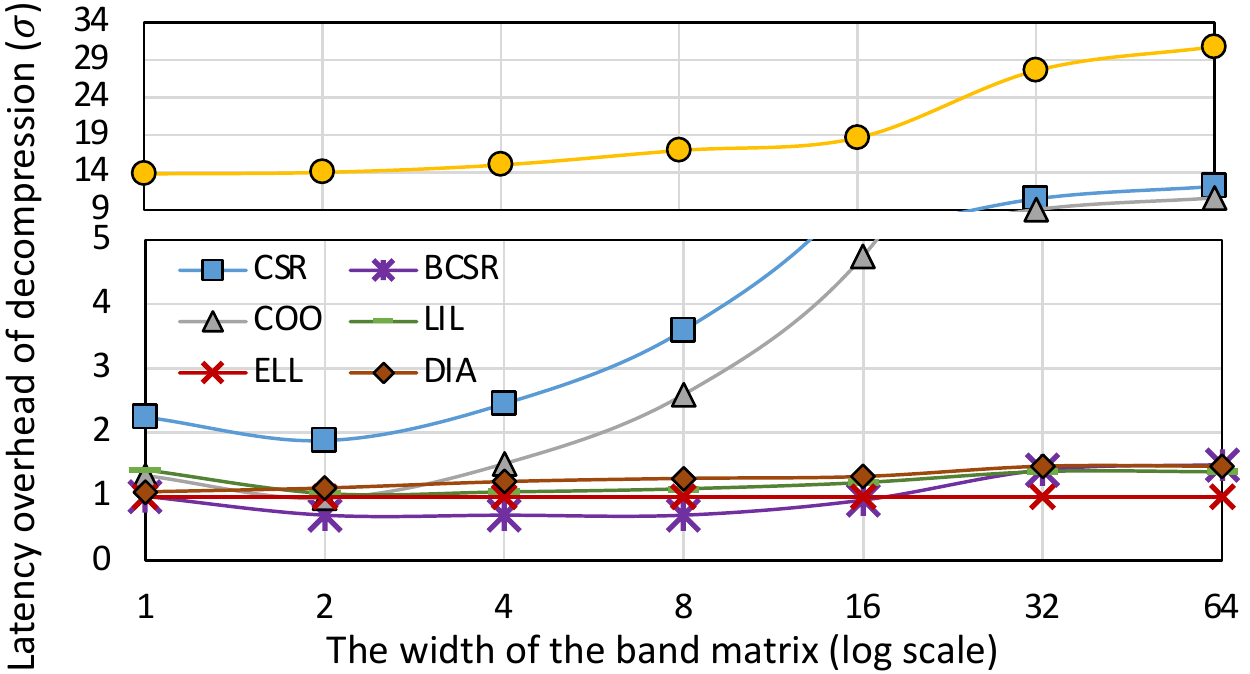}
\vspace{-10pt}
\caption{\textit{Decompression overhead for band matrices:} comparing the latency overhead, $\sigma$, of seven sparse formats for partition size of $16 \times 16$ when the width varies from 1 to 64.}
\vspace{-0pt}
\label{fig:latency_band}
\end{figure}

\setcounter{figure}{7}
\begin{figure*}[b]
\centering
\vspace{-10pt}
\includegraphics[width=1\textwidth]{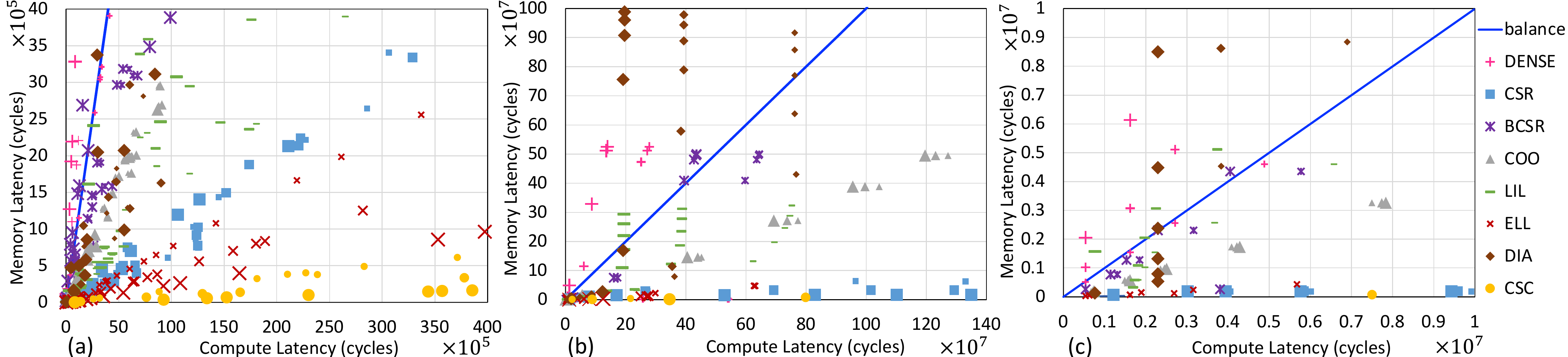}
\vspace{-20pt}
\caption{\textit{Balance ratio:} the relationship between the memory and compute latency for various partition sizes indicated by the size of markers for (a) SuiteSparse, (b) random workloads, and (c) band matrices. The blue line indicates balance ratio = 1.}
\vspace{-25pt}
\label{fig:balance_SS}
\end{figure*}

\setcounter{figure}{6}
\begin{figure}[t]
\centering
\vspace{-0pt}
\includegraphics[width=0.45\textwidth]{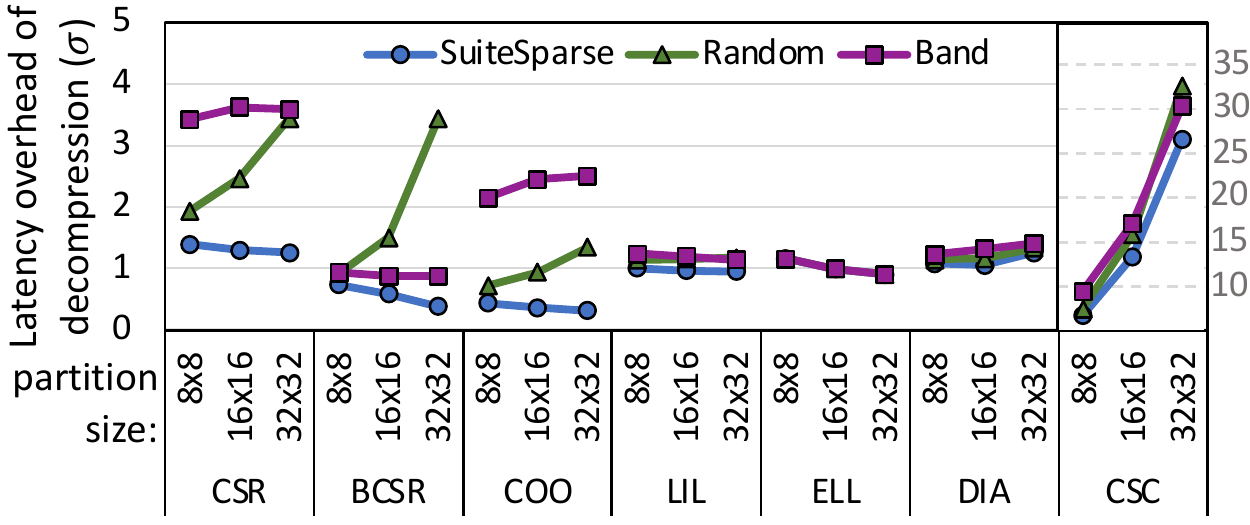}
\vspace{-10pt}
\caption{\textit{Decompression overhead for various partition sizes:} comparing the average $\sigma$ (lower is better) of seven sparse formats for three types of workloads (SuiteSparse, random, band) and partition sizes of 8, 16, and 32.}
\vspace{-18pt}
\label{fig:latency_part}
\end{figure}

This section explores $\sigma$, the latency overhead of decompression (Figure~\ref{fig:micro}, \ding{182}) for SuiteSparse, random, and structured band matrices in Figures~\ref{fig:latency_SS}, \ref{fig:latency_rand}, and \ref{fig:latency_band}, respectively. In Figures~\ref{fig:latency_SS}, the bars lower than one illustrate faster computation than the baseline dense format. The overhead of the dense baseline, for which $\sigma = 1$, is computing and transferring zero eateries, and the overhead of all sparse formats (only for computations) goes to the decompression. As Figures~\ref{fig:latency_SS} shows, the overhead of sparse formats can, in some cases, exceed that of the dense format. The worst-case scenario of decompression occurs with the CSC format because the orientation of data is opposite to that of the computation mechanism in hardware. 

From Figures~\ref{fig:latency_SS}, we do not observes any relationship between the density (darkness of the bars in Figures~\ref{fig:latency_SS}) and $\sigma$ in highly sparse matrices. Thus, Figures~\ref{fig:latency_rand} clarifies such a relationship for a wider range of density based on our randomly generated synthetic workloads. Likewise, Figure~\ref{fig:latency_band} shows the latency of band (and diagonal) matrices when the width increases. As the two figures illustrate, although the $\sigma$ of all formats increase with density and width of band matrices, it more dramatically increases for COO, CSR, and CSC. Besides, the time to reconstruct the rows of a matrix from a column-oriented compression format (i.e., CSC) leads to up to 21$\times$ and 30$\times$ slower computation than if we were to process all zero entries of the dense format, respectively, for random and band matrices. In such cases, preprocessing the sparse data to a format compatible with a hardware accelerator is highly suggested. 

Figures~\ref{fig:latency_part} illustrates the impact of partition size on $\sigma$. In all workloads (i.e., SuiteSparse, random, and structured), the computation latency of ELL is proportional to that of the dense format and does not change with the pattern of sparsity. This is because, in ELL, we are still processing a whole non-zero matrix regardless of its individual entries. However, since the length of these new squares (in our case, six) is smaller than that of the original dense partitions (8, 16, or 32), the computation latency of ELL decreases as the partition size increases. Seeking a relatively generic sparse format that can provide moderate computation latency for random and structured matrices, BCSR could be a fair option. However, it is not as good for random matrices when the partition size increases. This is because of the additional dot products that must be done per each non-zero block regardless of the individual values of the entries.

\setcounter{figure}{8}
\begin{figure*}[t]
\centering
\vspace{-15pt}
\includegraphics[width=1\textwidth]{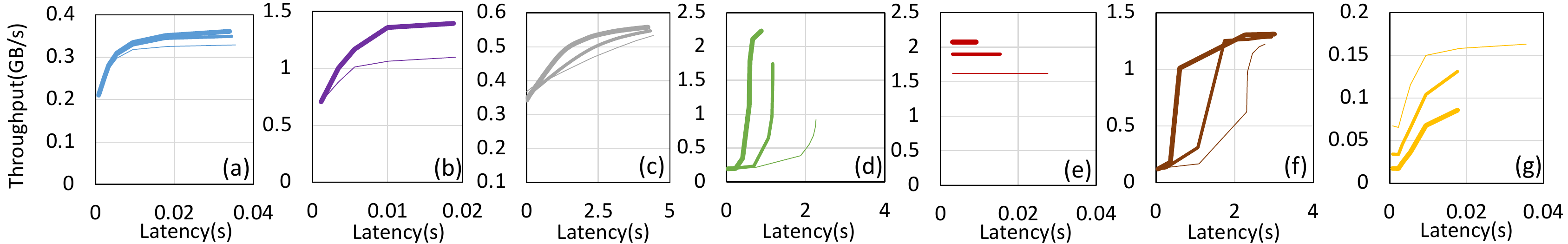}
\vspace{-20pt}
\caption{\textit{Throughput vs. latency:} comparing the throughput and the average time to apply SpMV on an $8000 \times 8000$: (a) CSR, (b) BCSR, (c) COO, (d) LIL, (e) ELL, (f) DIA, and (g) CSC. Thicker lines indicate larger partition sizes. }
\vspace{-12pt}
\label{fig:throughput_latency}
\end{figure*}


\subsection{Latency \& Balance Ratio}
\label{sec:balance}

Since memory accesses and computation are pipelined, the sum of their \textit{maximum for each partition} defines the total latency. Therefore, the latency overhead (discussed in Section~\ref{sec:latency}), which stems from only the computation, does not provide any information about which one (computation or memory) defines the total latency. Details in that regard are discussed in this section. Figure~\ref{fig:balance_SS} shows both memory and compute latency and thus implicitly shows the balance ratio: points below the \textit{balance} line have balance ratio < 1. 

Since only non-zero entries of non-zero partitions are transmitted, the latency to transmit data and metadata (i.e., memory latency) for all sparse formats is much lower than that for the dense format, as expected. The computation latency of sparse formats, however, is not always lower than for the dense format. While some of the frequently used formats such as CSR, CSC, and DIA have been successful in lowering the memory latency, their computation latency is higher than the baseline, which negates their benefits. 
%
Other formats such as LIL and ELL also do well in reducing memory latency (i.e., data-transfer time) but have similar computations as the baseline. For instance, the computation latency of LIL is defined by the longest column; hence, in some cases, it is the same as or more than that of dense format. For ELL, on the other hand, when the width of the ELL matrix is slightly smaller than the width of the original partition (e.g., the 8$\times$8 case), the computation latency of ELL is just slightly higher than dense format because of the overhead of decompression, even though it is small. Similar to random and structured matrices, the CSC format is the slowest for SuitSparse workloads with up to 27$\times$ higher latency compared to the dense baseline. All in all, Figure~\ref{fig:balance_SS} suggests that in terms of latency, COO or BCSR could be appropriate candidates (even though in some cases they perform as good as the dense format) to be used for diverse matrices from scientific and graph applications.

As Figure~\ref{fig:balance_SS} shows, for all types of matrices (i.e., SuiteSparse, random, and band), the balance ratio of dense format is higher than most of the sparse formats. This is because the zero entries impact both memory and computation latency. In fact, the balance ratio of dense format is closer to one (the perfect case) -- but it moves toward a memory-bound as partition size (indicated by marker size) increases. In some formats, such as LIL, increasing the partition size helps to achieve a better balance, while in some others, such as ELL, it is the opposite.
In random and structured band sparse matrices (Figures~\ref{fig:balance_SS}b and \ref{fig:balance_SS}c), higher density and/or larger bandwidth of band matrices leads to a memory bottleneck for BCSR, LIL, and DIA. In such cases, if adding more memory bandwidth to the system is possible, using BCSR or LIL for less sparse application (e.g., for the inference of neural networks) or using DIA for applications with diagonal/band matrices is suggested. Otherwise, COO seems to offer a reasonable balance for various densities as well as the varieties of band matrices. The same hypotheses for balance ratio are also applicable for the more diverse SuiteSparse workloads, as shown in Figure~\ref{fig:balance_SS}a.

\subsection{Throughput \& Bandwidth Utilization}
\label{sec:throughput_bw}

This section studies throughput and memory bandwidth utilization. First, Figure~\ref{fig:throughput_latency} explores the relationship between throughput and the total time to process an $8000 \times 8000$ matrix. The following parameters contribute to throughput: (i) the total processed data consisting of data and metadata and (ii) the total time to process, which is the maximum of memory latency (data-transfer time) and computation latency for each partition. As a result, in a sparse format such as ELL in which both total latency and data grow with the same pace, throughput does not change with latency (this is also the case for the dense baseline). For all formats but ELL, throughput increases with latency and then reaches a maximum. As Figure~\ref{fig:throughput_latency} suggests, BCSR, LIL, and DIA reach a higher throughput compared to the other four formats. Besides, for all formats but CSC, increasing partition size, shown by the thickness of lines in Figure~\ref{fig:throughput_latency}, results in higher throughput because both latency and data decrease as the partition size increases. Since throughput does not reflect the impact of transmitting and computing \textit{useful} data, we study throughput along with the utilization of memory bandwidth.

\begin{figure}[t]
\centering
\vspace{-0pt}
\includegraphics[width=0.45\textwidth]{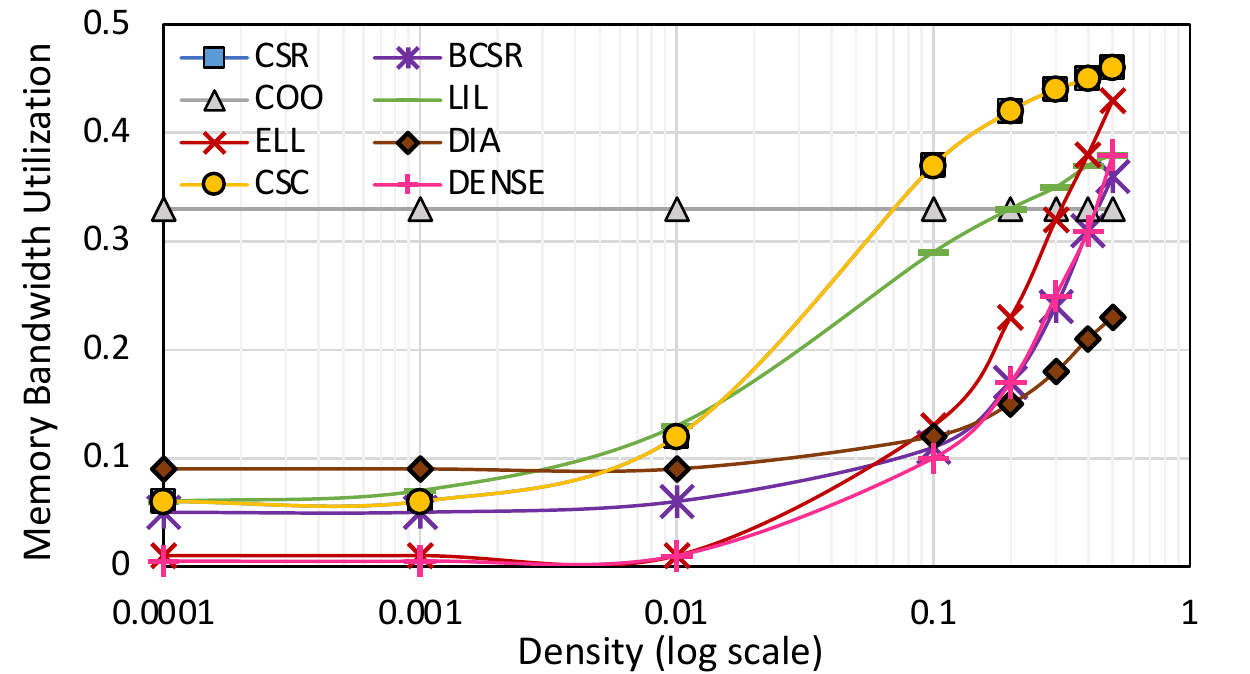}
\vspace{-10pt}
\caption{\textit{Memory bandwidth utilization for random matrices:} comparing seven sparse formats for partition size of $16 \times 16$ when density varies from 0.0001 to 0.5.}
\vspace{-18pt}
\label{fig:BW_rand}
\end{figure}

\begin{figure}[b]
\centering
\vspace{-18pt}
\includegraphics[width=0.45\textwidth]{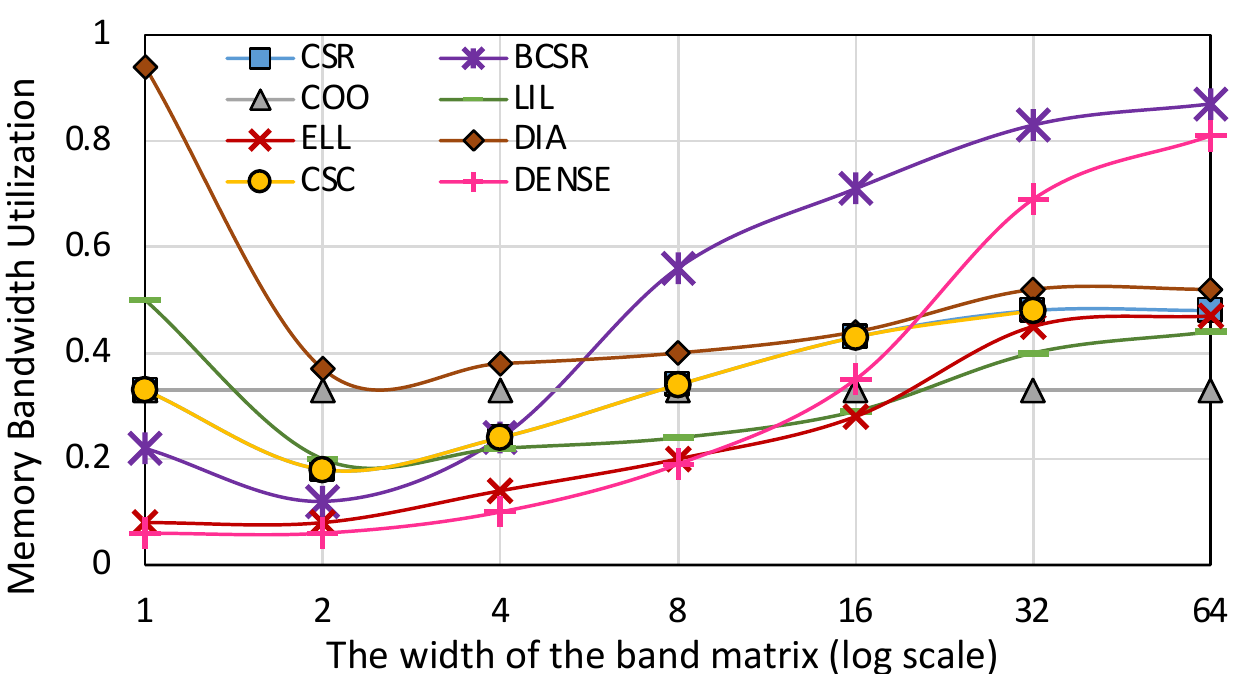}
\vspace{-10pt}
\caption{\textit{Memory bandwidth utilization for band matrices:} comparing seven sparse formats for partition size of $16 \times 16$ when the width varies from 1 to 64.}
\vspace{-0pt}
\label{fig:BW_band}
\end{figure}

\begin{figure}[t]
\centering
\vspace{-0pt}
\includegraphics[width=0.45\textwidth]{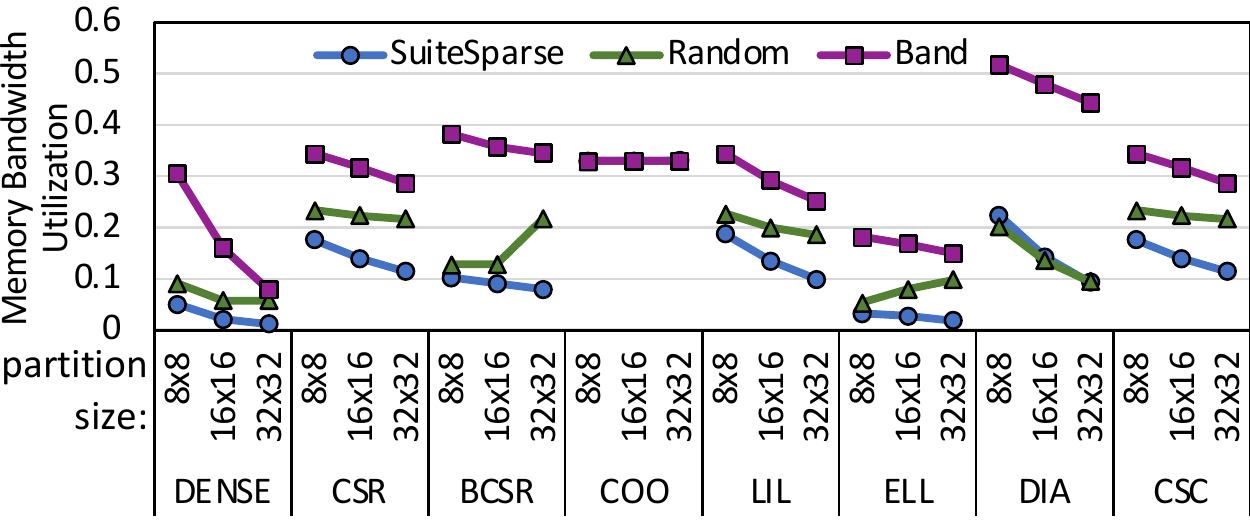}
\vspace{-10pt}
\caption{\textit{Memory bandwidth utilization for various partition sizes:} comparing the average memory bandwidth utilization (higher is better) on SuiteSparse, random, band workloads for partition sizes of 8, 16, and 32.}
\vspace{-20pt}
\label{fig:BW_part}
\end{figure}

Memory bandwidth utilization and its relationship with density, width of band matrices, and partition size are shown in Figures~\ref{fig:BW_rand},\ref{fig:BW_band}, and \ref{fig:BW_part}, respectively. As they all indicate, the memory bandwidth utilization of COO is always 0.3 since it always transmits two indices per one non-zero entry. Besides, as Figure~\ref{fig:BW_band} indicates, the memory bandwidth utilization of DIA for diagonal matrices is close to one -- the slight difference occurs because of saving the diagonal number for the main diagonal. As partition size grows, this memory bandwidth utilization approaches full utilization. However, for other band matrices, we see that the DIA format does not offer better memory bandwidth compared to \textit{more generic} formats such as COO, ELL, or LIL, among which LIL is a better candidate to cover more extreme sparseness as well as a wider variety of random matrices (Figure~\ref{fig:BW_rand}) while offering a better balance ratio at larger partitions compared to COO and ELL. Finally, Figure~\ref{fig:BW_part} demonstrates that, as expected, for all formats but COO, the memory bandwidth utilization of denser matrices (density $>$ 0.1) and structured ones is higher than that of extremely sparse matrices (e.g., SuiteSparse).


\subsection{Resource Utilization \& Power Consumption}
\label{sec:util}

Table~\ref{table:util} compares FPGA resource utilization and the dynamic power consumption. We must dedicate enough BRAM blocks to envision the worst-case scenarios even though they occur rarely. The other factor impacting the BRAM utilization is the degree of parallelism. To enable parallelism, we partition the matrices and distribute them to BRAM blocks. Because of these two factors, we see that CSR and CSC utilized the lowest number of BRAM blocks, whereas BCSR utilizes the same blocks as the dense implementation does. In some cases, such as ELL, smaller partitions (i.e., 8 and 16) use more flip flops (FFs) compared to larger partitions (i.e., 32). This is because, in a small partition size, the buffering is automatically implemented using FFs rather than BRAM blocks. It is also demonstrated by the fewer BRAM blocks utilized by the 8$\times$8 ELL. 

\begin{figure}[b]
\centering
\vspace{-15pt}
\includegraphics[width=0.5\textwidth]{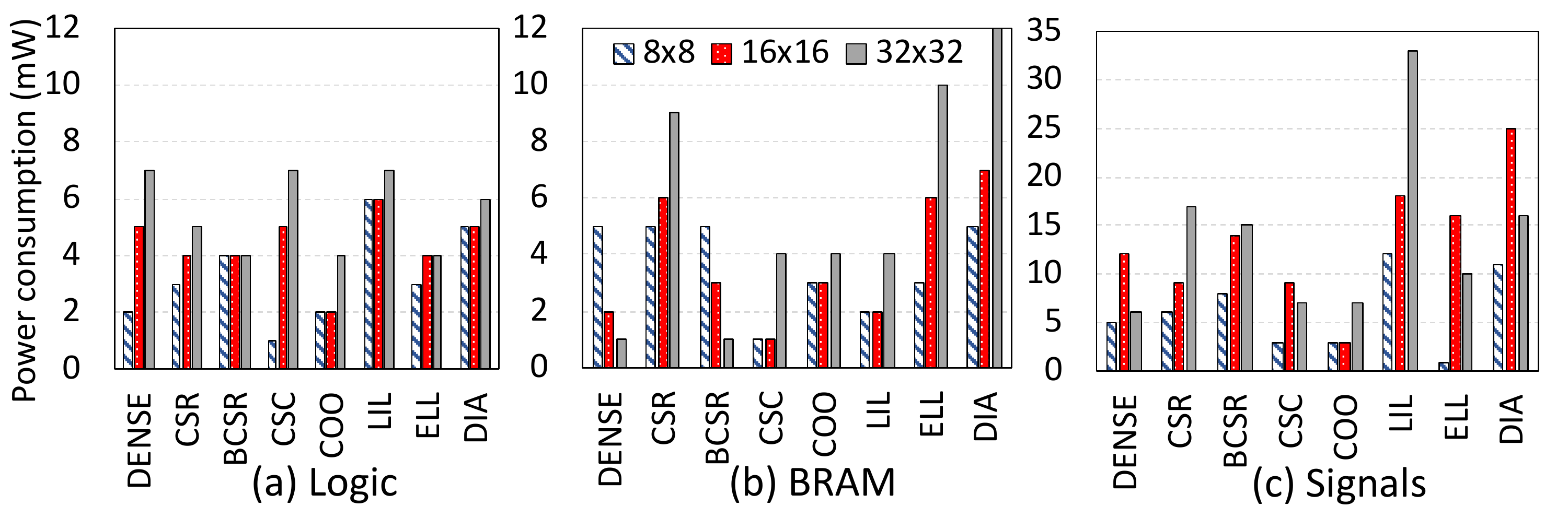}
\vspace{-20pt}
\caption{\textit{Dynamic power consumption:} breakdown into (a) logic, \textbf{(b)} BRAM, and \textbf{(c)} signals.}
\vspace{-0pt}
\label{fig:power}
\end{figure}

The dynamic power consumption listed in Table~\ref{table:util} suggests that while larger partition sizes cause higher power consumption in some formats (i.e., CSR, BCSR, COO, and LIL), for the others (i.e., dense, CSC, ELL, and DIA), the maximum power is consumed at the 16$\times$16 partition size and the minimum case may occur at 8$\times$8 (e.g., dense and CSC) or at 32$\times$32 (e.g., ELL and DIA). To clarify, Figure~\ref{fig:power}a, \ref{fig:power}b, and \ref{fig:power}c illustrate the dynamic power consumed by logic, BRAM, and signals, respectively. As Figure~\ref{fig:power} shows, the power consumption of logic always increases or stays steady as partition size increases, while that of BRAM may decrease (e.g., dense and BCSR). Therefore, comparing Figure~\ref{fig:power} against the dynamic power listed in Table~\ref{table:util} indicates that the trend of overall dynamic power consumption partially depends on BRAM, but more generally follows the same trend as the power consumption of signals (Figure~\ref{fig:power}c). By evaluating total latency and power consumption together, we see that for SuitSparse matrices, not only does COO consume the least dynamic power, but also it is the fastest in terms of total latency. However, if achieving high throughput at lower power is the goal, BCSR is a better fit. On the other hand, for structured matrices, LIL and ELL are the fastest in terms of latency and throughput, among which ELL performs better for band matrices with wider bandwidths and consumes less power. The static power consumption of dense, CSR, BCSR, LIL, and ELL is 0.121W and that of CSC, COO, and DIA is 0.103W. The static energy, which depends on time, can be an issue for those slower sparse formats that require less amount of dynamic energy.

\renewcommand{\arraystretch}{0.9}
\begin {table}[t]
\small
\begin{center} 
\vspace{-0pt}
\caption{Resource utilization and the total dynamic power consumption for three partition sizes (8, 16, and 32).}{
\vspace{-10pt}
\resizebox{1\columnwidth}{!}{
\begin{tabular}{c | c c c | c c c | c c c | c c c }
              \toprule
              & \multicolumn{3}{c|}{\textbf{BRAM\_18K}} & \multicolumn{3}{c|}{\textbf{FF ($\times$1000)}} & \multicolumn{3}{c|}{\textbf{LUT ($\times$1000)}} & \multicolumn{3}{c}{\textbf{DY Power(W)}}  \\
              
              \textbf{part. size:} & \textbf{8} & \textbf{16} & \textbf{32} & \textbf{8} & \textbf{16} & \textbf{32} & \textbf{8} & \textbf{16} & \textbf{32} & \textbf{8} & \textbf{16} & \textbf{32} \\  
              \midrule
             \textbf{DENSE} & 8 & 16 & 32 &	1.5 & 1.9 & 4.3 & 0.7 & 0.7 & 1.2 & 0.02 & 0.08 & 0.03\\
             \textbf{CSR} & 2 &	2 &	8 &	0.7	& 0.8 &	3.8 &	0.9 & 0.9 & 1.1 & 0.04 & 0.04 & 0.07 \\
             \textbf{BCSR} & 8 & 16 & 32 & 1.6 & 2.4 & 4.4 & 1.2 & 1.4 &	2.2 & 0.05 & 0.06 &	0.06 \\
             \textbf{CSC} & 1 &	1 &	9 &	0.9 & 1 & 2.7 &	1 &	1.2 &	1.1 & 0.01 & 0.05 & 0.03 \\
             \textbf{LIL} & 4 &	4 &	6 &	2.9 & 5.8 & 9.1 & 1.6 &	2.7 & 4.8 & 0.05 & 0.08 & 0.07 \\
             \textbf{ELL} & 1 &	7 &	9 &	2 & 3.2 & 0.9 &	0.9 & 1 &	0.8 & 0.06 & 0.10 & 0.06\\
             \textbf{COO} & 3 &	3 &	8 & 1.8 &	1.3 & 3.2	& 1.2 & 2.5 & 5.4 & 0.02 & 0.04 & 0.04 \\
             \textbf{DIA} & 3 &	3 &	11 & 2.2 &	5 & 9.2 & 1.5 &	2.8 & 4.6 & 0.07 & 0.12 & 0.05 \\
              \midrule
             \textbf{Total} & \multicolumn{3}{c|}{140} & \multicolumn{3}{c|}{106.4} & \multicolumn{3}{c|}{53.2} & \multicolumn{3}{c}{N/A}  \\
             \bottomrule
\end{tabular}}}
\label{table:util}
\end{center}
\vspace{-20pt}
\end{table}
\renewcommand{\arraystretch}{1.0}

%% file: TEX/related.tex
Proposing DSAs for SpMV in various application domains have been the topic of several recent studies~\cite{qin2020sigma, zhang2020sparch, srivastava2020tensaurus, gondimalla2019sparten, kanellopoulos2019smash, sadi2019efficient, hegde2019extensor, asgari2020alrescha}, in which sparse formats play a key role. While SIGMA~\cite{qin2020sigma}, Two-Step~\cite{sadi2019efficient}, and ExTensor~\cite{hegde2019extensor} use popular sparse formats such as CSR, CSC, and COO, others tailor a sparse format to their DSA~\cite{zhang2020sparch, asgari2020alrescha} or propose new formats~\cite{srivastava2020tensaurus, gondimalla2019sparten, kanellopoulos2019smash} as follows: SpArch~\cite{zhang2020sparch} proposes a condensed format, which is equivalent to storing a matrix in the CSR format and fetching the elements with the same index for all rows; Alrescha~\cite{asgari2020alrescha} modifies BCSR so that data follow a desired order dictated by DSA; Tensaurus~\cite{srivastava2020tensaurus} proposes compressed interleaved sparse slice (CISS) format, which allows accessing sparse data in a vectorized and streaming manner; SparTen~\cite{gondimalla2019sparten} suggests a bitmask representation, in which a sparse tensor is a two tuple of a bit mask, called SparseMap, and a set of non-zero values; and SMASH~\cite{kanellopoulos2019smash} presents a hierarchical bitmap compression that uses a non-zero values array (NZA) for holding the values.

%% file: TEX/insights.tex
\begin{wrapfigure}{r}{0.25\textwidth}
  \begin{center}
  \vspace{-25pt}
    \hspace{-20pt}\includegraphics[width=0.22\textwidth]{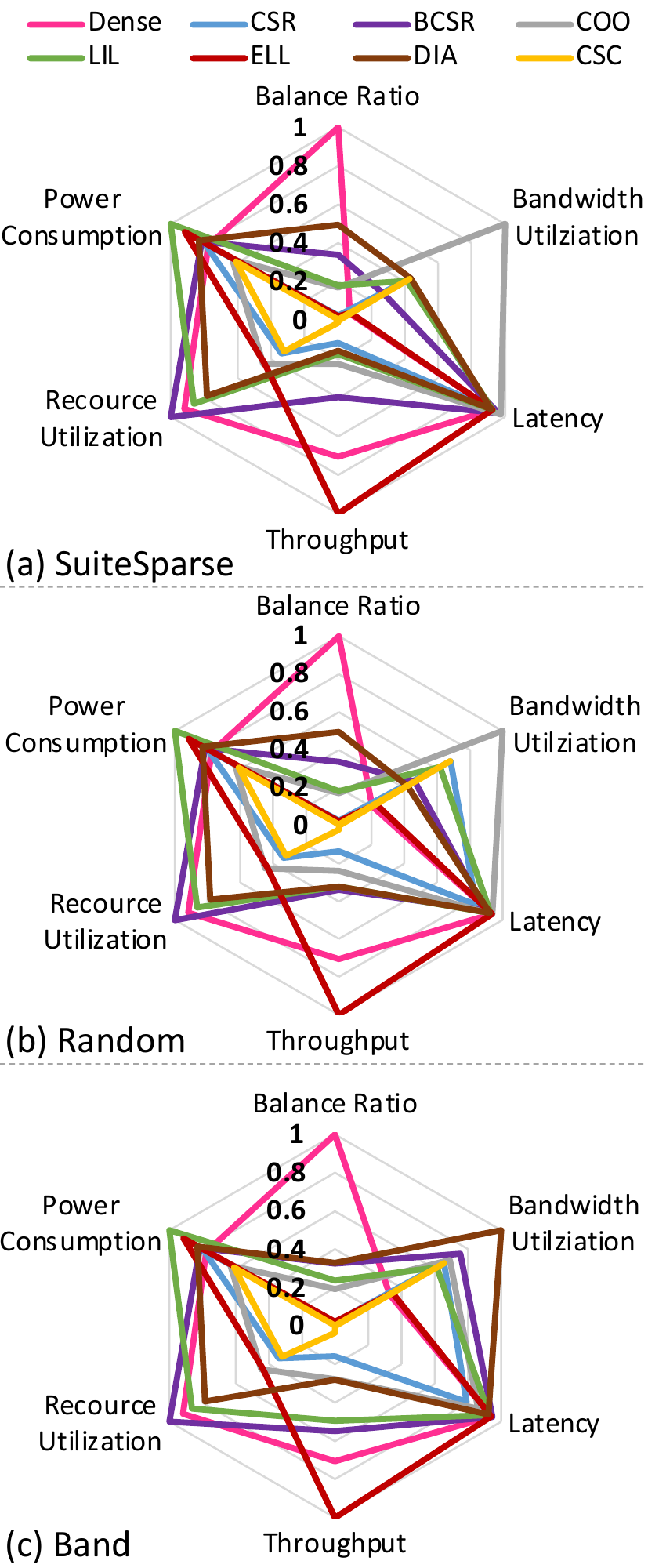}
  \end{center}
  \vspace{-10pt}
   \caption{{\small Comparing sparse formats for: (a) SuiteSparse, (b) random, and (c) band matrices. $1$ and $0$ show best and worst, respectively. }}
  \vspace{-13pt}
  \label{fig:insights}
\end{wrapfigure}

Copernicus investigated the performance implication of HLS-based FPGA implementations of seven sparse formats. We hope our results lead architects to knowingly choose the required sparse format and tailor their FPGA designs for sparse applications. Figure~\ref{fig:insights} summarizes all the six metrics for three group of workloads by normalizing each metric to its maximum achieved number so that "1" represents the best case and "0" represents the worst case. Here, we overview some of the key insights:

\noindent\textbullet\ Unlike a common belief, the memory bandwidth is not always the bottleneck; hence the performance sparse problems cannot always be improved by simply adding more memory bandwidth to the system. Thus, when using a format such as CSR to efficiently use storage, a lower-bandwidth low-cost memory is sufficient. Otherwise, the implementation of the computations must be further improved (if possible).

 \noindent\textbullet\ Although in scientific computation and graph analytics (Figure~\ref{fig:insights}a), the common patterns of sparse matrices are diagonal and band, our study shows that a non-specialized format such as COO performs faster and better utilizes the memory bandwidth compared to a specialized format such as DIA. This is because of the compatibility of more generic formats with a generic hardware for common computations. Besides, a generic format better tolerate the variations in the distribution of non-zero entries. If power consumption and FPGA resource utilization must also be considered, LIL or BCSR are other candidates.

\noindent\textbullet\ For structured band matrices (Figure~\ref{fig:insights}c), a pattern-specific format such as DIA, near-perfectly utilizes the memory bandwidth and does it better as the partition size increases. However, to allow such utilization to effectively impact the other performance metrics, the computation engine must also be tailored to the format if DIA must be used in a particular application. Otherwise, the mismatch would create a computation bottleneck.

\noindent\textbullet\ For less sparse (density $>$ 0.1) applications such as the inference of neural network, optimizations beyond simple partitioning of size 8$\times$8 or at most 16$\times$16 hurt the performance even though it might help reduce the memory footprint (possibly, not too much). Extracting the non-zero partitions from the neural network can be done with the aid of structure pruning schemes~\cite{wen2016learning, mao2017exploring, molchanov2016pruning, kung2019packing, asgari2019eridanus}.